\documentclass[sigconf]{acmart}
\usepackage{balance}
\AtBeginDocument{%
  \providecommand\BibTeX{{%
    \normalfont B\kern-0.5em{\scshape i\kern-0.25em b}\kern-0.8em\TeX}}}

\begin{document}
% \begin{sloppypar}
% interwordspace: \the\fontdimen2\font \\
% interwordstretch: \the\fontdimen3\font \\
% emergencystretch: \the\emergencystretch\par
% \blindtext
% \end{sloppypar}

\showboxdepth=\maxdimen
\showboxbreadth=\maxdimen
\title{BuGL - A Cross-Language Dataset for Bug Localization}
\titlenote{\textbf{BuGL} is publicly available along with dataset, documentation and scripts at {\url{https://github.com/muvvasandeep/BuGL}}.
}

\author{Sandeep Muvva, A Eashaan Rao, Sridhar Chimalakonda}
\affiliation{Research in Intelligent Software \& Human Analytics (RISHA) Lab\\Indian Institute of Technology Tirupati India \\}
\email{{cs16b017,cs19s501,ch}@iittp.ac.in}

\renewcommand\footnotetextcopyrightpermission[1]{}
\settopmatter{printacmref=false}
\pagestyle{plain} 
\begin{abstract}
Bug Localization is the process of locating potential error-prone files or methods from a given bug report and source code. There is extensive research on bug localization in the literature that focuses on applying information retrieval techniques or machine learning/deep learning approaches or both, to detect location of bugs. The common premise for all approaches is the availability of a good dataset, which in this case, is the standard benchmark dataset that comprises of 6 Java projects and in some cases, more than 6 Java projects.
The existing dataset do not comprise projects of other programming languages, despite of the need to investigate specific and cross project bug localization. To the best of our knowledge, we are not aware of any dataset that addresses this concern. In this paper, we present BuGL, a large-scale cross-language dataset. BuGL constitutes of more than 10,000 bug reports drawn from open-source projects written in four programming languages, namely C, C++, Java, and Python. The dataset consists of information which includes Bug Reports and Pull-Requests. BuGL aims to unfold new research opportunities in the area of bug localization.
\end{abstract}




\maketitle

\section{Introduction}
\label{intro}
Bug in a software system is an underlying cause of a fault/error which make the software to deviate from its correct behaviour \cite{wong2016survey}. Bug Localization task is to locate the bug position in the source code and is considered as one of the critical activities in software maintenance cycle. During pre and post-release, many bugs are identified and reported \cite{wong2016survey}. Automatically identifying the most probable source that causes the defect could greatly reduce the effort of developers in localizing the source of the bug \cite{polisetty2019usefulness}. Bug description comes in the form of bug reports that majorly contain natural text \cite{koyuncu2019d}. It also accommodates stack traces and code snippets, that provide structured information related to a bug \cite{rath2019structured}. For locating the source of the bug using bug reports, two prominent strategies - Information Retrieval (IR) and Hybrid approach consisting of IR + Machine Learning (ML) + Deep Learning (DL) have been suggested by researchers.\

IR addresses Bug Localization as a "document retrieval problem" \cite{chaparro2019using}. From the textual information present in the bug report, queries are fired to fetch top-ranked source files that could possibly contain bugs. Many IR based techniques and tools exist in the literature \cite{lukins2010bug, saha2013improving, zhou2012should, youm2015bug}. Lukins et al. \cite{lukins2010bug}, created a static LDA model from the source code that generates n-topics having n-words each. This model can be queried for bugs that result in a set of files, ranked based on the probability distribution. The projects used for this technique are Mozilla, Eclipse, and Rhino. \textit{BugLocator}, utilizes the revised Vector Space Model (rVSM) \cite{zhou2012should} to rank files based on the textual similarity between source code and bug report, along with the history of similar bug fixes taken from Eclipse, AspectJ, SWT, and Zxing. Similar projects have been tested by \textit{BLUiR} tool, that employs class names, method names, and bug summaries to rank files and was observed that these rankings outperform BugLocator \cite{saha2013improving}. Some other criteria, such as version history \cite{youm2015bug} and similar bug history \cite{youm2015bug, wang2014version}, have been considered, along with bug reports and source code for bug localization. Certain techniques such as segmentation \cite{wong2014boosting}, where source code is bifurcated into segments, and stack trace, which contains all the data related to invocation calls till an exception is encountered \cite{wong2014boosting, rath2019structured} are studied and leveraged in bug localization due to the structured information they provide. Stack-trace analysis and segmentation are complementary to each other and thus provide a better ranking of files \cite{wong2014boosting}. Despite the availability of many techniques, current state-of-the-art IR-based tools are still observed to be unreliable \cite{le2017will}.

Hybrid approaches consist of a mixture of methods (IR, ML, and DL) to reduce the "lexical gap" that exists between source code and bug reports and provides higher accuracy in bug localization \cite{polisetty2019usefulness}. Koyuncu et al. \cite{koyuncu2019d} have proposed a divide and conquer IR based approach and also examined the query formulation and its impacts on the localization performance. They implemented a multi-classifier approach to compute weights and assign them to the features extracted from bug reports and source code. A training dataset which contains exact bug-location pairs has been created and was passed to the gradient-boosting method to build multiple classifier model. \textit{DNNLOC} \cite{lam2017bug} leverages the features extracted from textual similarity of a bug report and source code using rVSM. DNN learns these extracted features and relates them with code tokens from the source code \cite{lam2017bug}. \textit{DeepLocator}  \cite{xiao2017improving} uses enhanced CNN for utilizing full semantic information and bug-fixing history, which are available in AspectJ, Eclipse, JDT, SWT and Tomcat projects. It uses \textit{revised} Term Frequency-Inverse Document Frequency (rTF-IDuF) to find relevant terms from the bug reports. Word embedding technique, i.e., word2vec, converts relevant terms from bug reports and AST from source code into vectors. These vectors are passed on to CNN, which predicts the localized file for a given bug. In view of cross-project bug localization, \textit{TRANP-CNN} \cite{huo2019deep} extracts semantic features available from the source project and maps it to the target project. A system named CAST, which extracts lexical and semantic information from the bug report and source code, applies tree-based CNN along with customized AST \cite{liang2019deep}.

A common point for all these techniques is that the projects on which these techniques are applied are written in Java. \textbf{AspectJ, Eclipse, JDT, SWT, Tomcat, Zxing} are the most commonly used projects in bug localization \cite{polisetty2019usefulness, liang2019deep, xiao2017improving, lam2017bug, wong2014boosting, youm2015bug, wang2014version, saha2013improving, zhou2012should}. While Rath et al. have used ``IlmSeven'' \cite{rath2017ilmseven} dataset in \cite{rath2019structured}, TRANP-CNN used dataset containing 3 Java projects - HTTPClient, Jackrabbit, Lucene-Java \cite{kochhar2014potential}. Koyuncu et al. have used the benchmark dataset in bug localization known as Bench4BL \cite{lee2018bench4bl} in \cite{koyuncu2019d}.

% and \cite{koyuncu2019d} use 
% While Rath et al. \cite{rath2019structured} used  dataset  TRANP-CNN uses dataset contains 3 Java projects named as HTTPClient, Jackrabbit, Lucene-Java \cite{kochhar2014potential} and \cite{koyuncu2019d} use the benchmark dataset in bug localization known as Bench4BL \cite{lee2018bench4bl}.

Most of the bug localization approaches are applied only on few Java projects. These projects have bug data from well-known bug-tracking systems such as Bugzilla\footnote{https://www.bugzilla.org/} and JIRA\footnote{https://www.atlassian.com/software/jira}, and are available in proper format. However, there are many projects stored on open source code sharing platforms such as GitHub, where bug reports are not adequately documented or available. Thus, there is a need to expand the domain of bug localization with respect to bug reports availability and to improve the techniques to handle different projects written in multiple programming languages. This drive leads us to the creation of "BuGL" dataset. The dataset comprises of projects from four different programming languages- C, C++, Java, and Python. It consists of bug reports along with pull requests which fixed the issue mentioned in the bug report. BuGL collectively has more than 10K bugs. The projects chosen here are different from the ones mentioned in the literature. The purpose of creating this dataset is to allow researchers to undertake bug localization challenges which can have various characteristics such as distinct languages, bug reports, projects, techniques, and so on. 

There are few cross-language datasets available in the literature which provide a myriad of opportunities to understand the development process of a software, without being restricted to any particular programming language. GitHub is one of the largest sources for open software artifacts, being the source of extraction for many datasets including BuGL. \textit{Public Git Archive} \cite{markovtsev2018public} and \textit{GHTorrent} \cite{gousios2013ghtorent} are some of the large scale open-source code datasets extracted from GitHub. \textit{Software Heritage Graph dataset} is the largest "public archive of software source code that comes along with development history" \cite{pietri2019software}. 

Datasets have been proposed to study various aspects of software engineering in specific. Gousious et al. have proposed \textit{pullreqs} dataset to support the study of pull-based development models \cite{gousios2014dataset}. \textit{DupPR} has been created by Yu et al. to study duplicate pull requests \cite{yu2018dataset}. However, to the best of our knowledge, no cross-language dataset to study bug localization is available, motivating the creation of ``BuGL'' dataset.

% Some datasets were designed to study specific aspects of Software Engineering such as, the pull-based development model \cite{gousios2014dataset} proposed  dataset, while to study duplicate pull-requests \textit{DupPR} created \cite{yu2018dataset}. However, to the best of our knowledge, no cross-language dataset is available to study bug localization. It motivated us to create the dataset called ``BuGL''.

\section{Dataset}
\subsection{Objectives}
Currently, most of the bug localization techniques/tools are developed based on existing benchmark dataset that consists only Java projects, which was constructed around 2014. Later in 2018, Lee et al. \cite{lee2018bench4bl} created a dataset of 51 projects (5 old subjects and 46 new subjects) and 10,017 bug reports from Java to conduct a reproducibility study of the performance of IR based bug localization techniques with a large number of subjects.  But whether these techniques/tools work as good as on the datasets constructed with other programming languages still remains as a question. This query motivated us to construct BuGL, a dataset of four different programming languages, namely C, C++, Python, and Java, which could be used as a benchmark dataset for Bug Localization in the future along with comparative studies with the current benchmark dataset.

\subsection{Methodology for choosing the projects} \label{Criteria for projects}
GitHub is a good source of open source projects. From millions of public repositories, artifacts such as issues and pull requests are extracted and used for large-scale studies. GitHub has some features such as stars, best matches and so on, which infers the number of developers interested in the project, and thus acts as an index of popularity \cite{borges2018s}. For choosing the primary programming language of a project, we used GitHub's linguistic feature. Almost all the GitHub projects have open and closed issues and pull requests. We include projects having no less than 500 closed issues and pull requests, with the intuition that at least 100 issues could be correlated to pull requests which solved those issues, which, as a result could provide vital metadata information about bugs. 
%Being a good source of open source projects, we chose GitHub for constructing our dataset. We mainly focused on the active projects that most of the users are interested. GitHub feature, more number of stars \cite{borges2018s}, indicates that users are interested in that project. From these projects we selected those projects with a decent number of pull requests and issues.
\begin{figure}
    \centering
    \includegraphics[width=10cm, height=7cm]{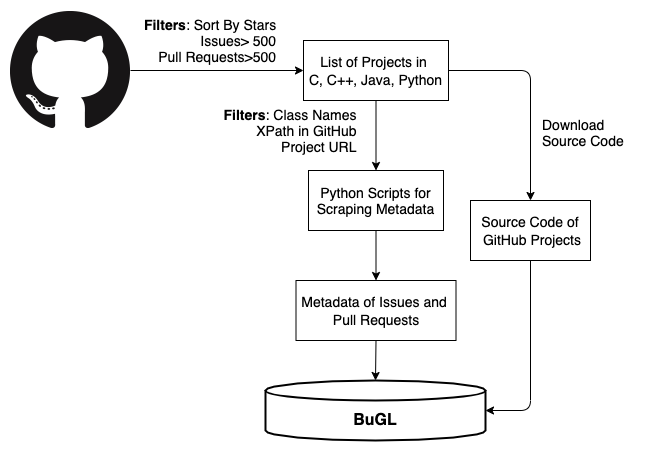}
    \caption{Methodology for curating the Dataset}
    \label{fig:methodology}
\end{figure}

\subsection{Methodology for filtering the bugs} \label{Criteria for Bugs}
Issues in GitHub repository usually do not have the important metadata such as files changed while resolving the issues, number of lines changed in each file, and so on, which play an essential role in Bug Localization, but, pull requests contain  such useful information. However, a pull request in GitHub need not be a bug fix, but could be an enhancement, bug, or a feature. Hence, pull requests cannot be considered as bugs by default. However, some of the assignees label a pull request with specific keywords such as bug, feature, and enhancement. But most of the pull requests do not have such labels. Thus, identifying correlation between pull requests and issues could fetch required metadata of the issues to be included in the dataset.

Issues have been correlated with pull requests based on the description of pull requests and keywords such as \textbf{fixes}, \textbf{resolves}, and so on, followed by \textbf{\#issue ID}. If the assignee merges a pull request after verification, then the issue is treated as closed. This implies that the merging of the pull request has resolved specific issue with the mentioned issue ID. The metadata from this pull request can be used for the issue raised in the repository.

% To correlate issues with pull requests, we have to check the description of the pull requests with keywords  If the assignee merges the pull request after verification, then that issue is treated as closed. It implies that the merging of the pull request resolves the issue with the mentioned issue ID. The metadata from this pull request can be used for the issue mentioned. 
\begin{figure}
    \centering
    \includegraphics[width=\linewidth, height = 4.5cm]{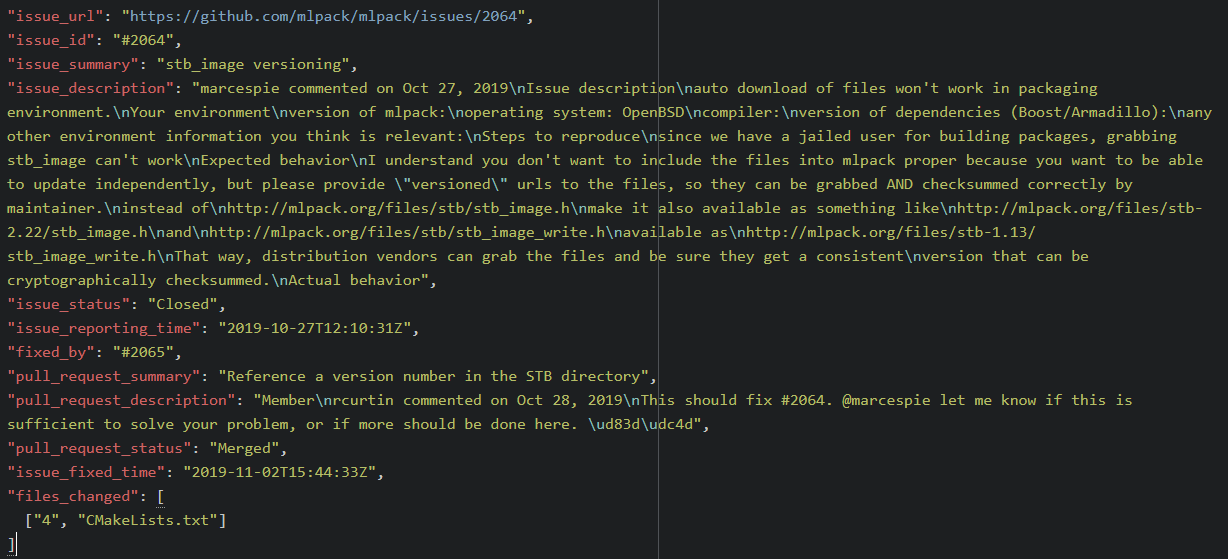}
    \caption{Metadata of a sample bug}
    \label{fig:sample bug}
\end{figure}

\subsection{Dataset Creation}
Projects present in the dataset are manually curated to carefully include only those projects that met the criteria mentioned in Section \ref{Criteria for projects}. Projects were extracted from four different languages- C, C++, Python, and Java, based on a three-step selection process, presented below:

\begin{itemize}
    \item Using GitHub's linguistic feature, projects were selected by identifying their primary programming language.
    \item Projects were sorted based on the number of stars.
    \item Manual selection of projects based on selection criteria mentioned in Section \ref{Criteria for projects}
\end{itemize}

Figure \ref{fig:methodology} summarises the methodology adopted for curating the dataset. We selected 54 projects consisting of 10,187 bugs, that included 21 projects from C, 11 from C++, 12 from Python, 10 from Java and downloaded all the selected projects. The aim is to accumulate at least 2000 issues from multiple projects for each programming language. That's why there are varying project numbers across these programming languages.

\begin{figure}
    \centering
    \includegraphics[width=\linewidth]{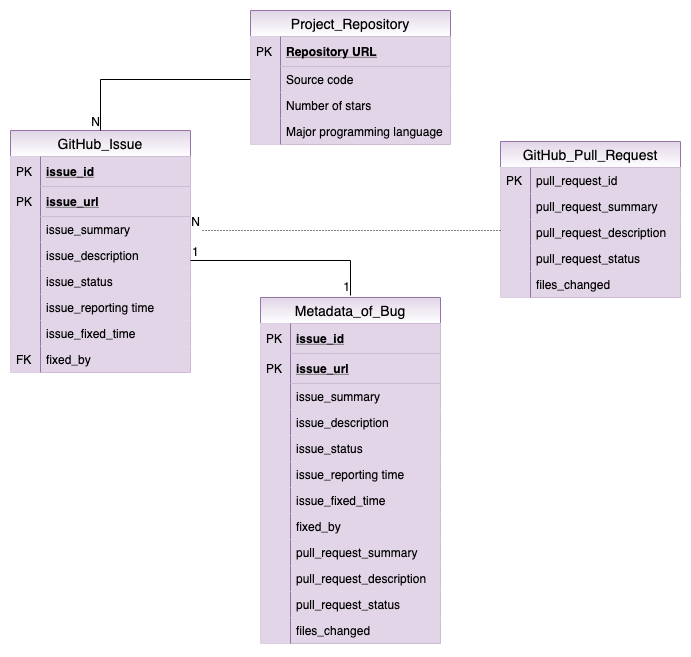}
    \caption{Database Schema}
    \label{fig:database schema}
\end{figure}

In the next step, we collected metadata of correlated pull requests and issues. We executed a python script and used selenium and chrome driver to scrape the required metadata and then, stored it in \textbf{json} and \textbf{xlsx} formats. We manually analyzed the GitHub page for each project and found some common CSS selectors like class name, id, or XPath, which were used to extract the metadata like issue id, description, and so on.  As shown in the Figure \ref{fig:sample bug}, metadata includes issue id, issue summary, issue description, issue reporting time, issue status, fixed by (id of the pull request which closes the issue on merging), pull request description, pull request status, files changed and number of files changed in each file. For open issues, metadata related to pull requests is empty. Figure \ref{fig:database schema} summarises the schema of BuGL, there are 4 tables namely \texttt{Project\_Repository, GitHub\_Pull\_Request, Metadata\_of\_Bug and GitHub\_Issue}. Here GitHub issues are mapped with GitHub Pull Request as mentioned in Section \ref{Criteria for Bugs}. Metadata for each bug can be found in the table \texttt{Metadata\_of\_Bug} having fields such as \texttt{issue\_id, issue\_summary, issue\_status, fixed\_by} and so on.

\subsection{Dataset Statistics}
Table \ref{tab:final stats} represents the statistics of BuGL and Table \ref{tab:projects and bugs stats} represents language-wise statistics of BuGL. The final dataset consists of metadata and source code of projects having at least 500 pull requests, and at the minimum 10 issues which have been resolved by pull requests. A \textbf{.xlsx} file (``BuGL details'' in the dataset) contains the name of the repository, number of stars, GitHub link, number of open and closed issues, number of open and closed pull requests, and number of bugs. No less than 10 projects from each programming language are included in the dataset.

\begin{table}[!htbp]
    \centering
    \begin{tabular}{|c c|}
    \hline
        Number of projects & 54 \\
        Number of closed issues & 151161 \\
        Average Number of issues & 2799 \\
        Total number of issues closed by pull requests & 10187 \\
        Average number of issues closed by pull requests & 189 \\
    \hline
    \end{tabular}
    \caption{Statistics of final dataset}
    \label{tab:final stats}
\end{table}

%Final dataset of BuGL consists of \textbf{x} number of projects from all the four languages C, C++, Python and Java with \textbf{y} number of bugs from all the projects. 

\begin{table*}[!htbp]
    \centering
    \begin{tabular}{|p{0.1\linewidth}|p{0.15\linewidth}|p{0.1\linewidth}|p{0.1\linewidth}|p{0.1\linewidth}|p{0.2\linewidth}|}
    \hline
        \textbf{Language} & \textbf{No. of Projects} & \textbf{Closed Pull Requests} & \textbf{Closed Issues} & \textbf{Open Issues} & \textbf{No. of Issues closed by Pull Requests}\\ 
        \hline
        C & 21 & 51408 & 36617 & 8608 & 2462\\
        C++ & 11 & 37198 & 30227 & 3607 & 2222\\
        Python & 12 & 32454 & 39760 & 6767 & 2626\\
        Java & 10 & 47258 & 44557 & 4210 & 2877\\
        \hline
    \end{tabular}
    \caption{Programming language wise statistics in BuGL}
    \label{tab:projects and bugs stats}
\end{table*}

%Here comes a paragraph describing the final stats of the dataset like how many issues for each language and a table showing the stats followed by a pictorial representation of how dataset has been collected.

%Our main motive in constructing BuGL was to construct a dataset which can be used as a benchmark dataset which consists of bugs from projects written mainly in four different programming languages namely C, C++, Java and Python , where current benchmark dataset being used consists of only Java projects which was constructed around 2014. Currently all the Bug Localisation techniques are developed based on this benchmark dataset which consists of only Java Projects. But will these techniques work on the projects written in other programming languages? 

%\section{Methodology}

\section{Research Opportunities}
BuGL could be a valuable resource in the area of Software Bugs. Our dataset has been carefully curated and provides attributes such as bug reports, pull requests, file changes, and other metadata with respect to bugs. The unique aspect of this dataset is the diversity of bugs from different programming languages, which makes it suitable for a wide range of use cases. The BuGL dataset could provide several research opportunities and could serve as a standard dataset for software-bug related studies.

Below we highlight a few research questions and insights that could be leveraged through BuGL.

\begin{itemize}
    \item \textbf{Bug Localization}: One of the main applications of our dataset is to facilitate research in locating source of the bug. It aids in comprehending bugs from the projects written in several programming languages. BuGL allows researchers to investigate topics relevant to bug localization such as \textit{Finding similarity between bugs occurring in different programming languages}, \textit{Testing if the existing bug-localization techniques work on BuGL}, \textit{Studying if change in programming languages affect bug localization process}, \textit{Analysing use of bug data on from projects in one programming language in facilitating bug localization for projects written in a different programming language}, \textit{Variation of bug localization techniques for projects developed in the open-source and proprietary environments}. We are working towards answering these questions using the proposed dataset.
    
    \item \textbf{Cross-Project Learning}: Another major usage of our dataset is to answer \textit{How effectively learning could take place from the data present in the source project, for an effective utilization of it in the target projects}. Studies done in this area have been observed to be scarce, and the data available in BuGL in the form of pull requests and bug reports can be utilized for cross-project learning. It is interesting to tackle some questions such as \textit{Possibility of cross-project learning of projects written in different programming languages in terms of continuous integration, defect prediction, software implementation, and so on.}
    
    \item \textbf{Open source development:} The software development community has embraced open-source philosophy, and it always serves the researcher to understand many aspects of Software building process. An exploration of bug reports and pull-requests triaging process, program comprehension, bug report quality analysis, software architecture in the context of bugs, and so on, could be performed using BuGL.
\end{itemize}

The scope of existing bug localization techniques is limited, as they worked and compared against a similar kind of projects mentioned in Section \ref{intro}. Hence, these methods might face an overfitting problem. The projects used in the approaches are well maintained and documented projects. However, many open-source projects are lack these features. Tackling these problems is essential, and the BuGL dataset is one of the first steps to deal with the problems mentioned above.

\section{Discussion}
Below, we list out a few scopes of improvement in the dataset along with key limitations of BuGL:
\begin{itemize}
    \item BuGL represents only a fraction of repositories from GitHub. However, in future, we plan to extend the dataset with more curated projects.
    \item Insufficient description of bug reports and pull requests makes it hard to categorize bug issues from other issues. Issues raised in the repository are sometimes are not bugs; these issues mostly deal with the documentation files or minor changes in the file name. To mitigate this, we mapped issues with the pull requests which resolved respective issues. Changes made in files by a pull request gives us information on whether the issues reported are related to the bug or not. 
    \item The projects selected from GitHub are based on number of stars. We applied, Borges and Valente \cite{borges2018s} recommendation while curating projects based on GitHub stars because popularity of some repositories might be due to their active promotion in the social media. 
    \item The issues present in projects are sometimes insufficient, and to train or test a model for bug-localization we need more issues. With this in mind, we included open issues in the dataset. It would help to enhance the model and might be useful for future upgradation of the dataset.
    \item To extend this dataset, we plan to include repositories from various domains and programming languages. The aim is to include a diverse range of bugs that can help to formulate new sets of bug localization techniques.
    \item More emphasis will be given towards adding new features in the dataset for more in-depth analysis of bug reports.
    \item We are also planning to create an automatic tool that could deal with duplicate bug reports and pull requests.
\end{itemize}

\section{Conclusion}
In this paper, we described BuGL, a cross-language dataset consisting of bug reports and pull request information. BuGL consists of more than 10k bug reports gathered from projects written in 4 different programming languages- C, C++, Java, and Python. We discussed the methodology used for constructing BuGL along with its representation. We also talk about the research opportunities related to this dataset. We hope BuGL creates a strong impact providing new directions and insights in Bug Localization. 

\section*{Acknolwedgements}
Even though cross project bug localization is a home-grown project, we would like to thank ARiSE Group, Robert Bosch Engineering and Business Solutions Ltd., Bangalore, India for funding some of our other work in bug localization.

\bibliographystyle{ACM-Reference-Format}
\balance{\bibliography{BuGL}}


\begin{thebibliography}{25}

%%% the Web address.

\ifx \showCODEN    \undefined \def \showCODEN     #1{\unskip}     \fi
\ifx \showDOI      \undefined \def \showDOI       #1{#1}\fi
\ifx \showISBNx    \undefined \def \showISBNx     #1{\unskip}     \fi
\ifx \showISBNxiii \undefined \def \showISBNxiii  #1{\unskip}     \fi
\ifx \showISSN     \undefined \def \showISSN      #1{\unskip}     \fi
\ifx \showLCCN     \undefined \def \showLCCN      #1{\unskip}     \fi
\ifx \shownote     \undefined \def \shownote      #1{#1}          \fi
\ifx \showarticletitle \undefined \def \showarticletitle #1{#1}   \fi
\ifx \showURL      \undefined \def \showURL       {\relax}        \fi
% invisible to TeX
\providecommand\bibfield[2]{#2}
\providecommand\bibinfo[2]{#2}
\providecommand\natexlab[1]{#1}
\providecommand\showeprint[2][]{arXiv:#2}

\bibitem[\protect\citeauthoryear{Borges and Valente}{Borges and
  Valente}{2018}]%
        {borges2018s}
\bibfield{author}{\bibinfo{person}{Hudson Borges} {and}
  \bibinfo{person}{Marco~Tulio Valente}.} \bibinfo{year}{2018}\natexlab{}.
\newblock \showarticletitle{What’s in a GitHub star? understanding repository
  starring practices in a social coding platform}.
\newblock \bibinfo{journal}{\emph{Journal of Systems and Software}}
  \bibinfo{volume}{146} (\bibinfo{year}{2018}), \bibinfo{pages}{112--129}.
\newblock


\bibitem[\protect\citeauthoryear{Chaparro, Florez, and Marcus}{Chaparro
  et~al\mbox{.}}{2019}]%
        {chaparro2019using}
\bibfield{author}{\bibinfo{person}{Oscar Chaparro},
  \bibinfo{person}{Juan~Manuel Florez}, {and} \bibinfo{person}{Andrian
  Marcus}.} \bibinfo{year}{2019}\natexlab{}.
\newblock \showarticletitle{Using bug descriptions to reformulate queries
  during text-retrieval-based bug localization}.
\newblock \bibinfo{journal}{\emph{Empirical Software Engineering}}
  \bibinfo{volume}{24}, \bibinfo{number}{5} (\bibinfo{year}{2019}),
  \bibinfo{pages}{2947--3007}.
\newblock


\bibitem[\protect\citeauthoryear{Gousios}{Gousios}{2013}]%
        {gousios2013ghtorent}
\bibfield{author}{\bibinfo{person}{Georgios Gousios}.}
  \bibinfo{year}{2013}\natexlab{}.
\newblock \showarticletitle{The GHTorent dataset and tool suite}. In
  \bibinfo{booktitle}{\emph{2013 10th Working Conference on Mining Software
  Repositories (MSR)}}. IEEE, \bibinfo{pages}{233--236}.
\newblock


\bibitem[\protect\citeauthoryear{Gousios and Zaidman}{Gousios and
  Zaidman}{2014}]%
        {gousios2014dataset}
\bibfield{author}{\bibinfo{person}{Georgios Gousios} {and}
  \bibinfo{person}{Andy Zaidman}.} \bibinfo{year}{2014}\natexlab{}.
\newblock \showarticletitle{A dataset for pull-based development research}. In
  \bibinfo{booktitle}{\emph{Proceedings of the 11th Working Conference on
  Mining Software Repositories}}. \bibinfo{pages}{368--371}.
\newblock


\bibitem[\protect\citeauthoryear{Huo, Thung, Li, Lo, and Shi}{Huo
  et~al\mbox{.}}{2019}]%
        {huo2019deep}
\bibfield{author}{\bibinfo{person}{Xuan Huo}, \bibinfo{person}{Ferdian Thung},
  \bibinfo{person}{Ming Li}, \bibinfo{person}{David Lo}, {and}
  \bibinfo{person}{Shu-Ting Shi}.} \bibinfo{year}{2019}\natexlab{}.
\newblock \showarticletitle{Deep transfer bug localization}.
\newblock \bibinfo{journal}{\emph{IEEE Transactions on Software Engineering}}
  (\bibinfo{year}{2019}).
\newblock


\bibitem[\protect\citeauthoryear{Kochhar, Tian, and Lo}{Kochhar
  et~al\mbox{.}}{2014}]%
        {kochhar2014potential}
\bibfield{author}{\bibinfo{person}{Pavneet~Singh Kochhar},
  \bibinfo{person}{Yuan Tian}, {and} \bibinfo{person}{David Lo}.}
  \bibinfo{year}{2014}\natexlab{}.
\newblock \showarticletitle{Potential biases in bug localization: Do they
  matter?}. In \bibinfo{booktitle}{\emph{Proceedings of the 29th ACM/IEEE
  international conference on Automated software engineering}}.
  \bibinfo{pages}{803--814}.
\newblock


\bibitem[\protect\citeauthoryear{Koyuncu, Bissyand{\'e}, Kim, Liu, Klein,
  Monperrus, and Traon}{Koyuncu et~al\mbox{.}}{2019}]%
        {koyuncu2019d}
\bibfield{author}{\bibinfo{person}{Anil Koyuncu},
  \bibinfo{person}{Tegawend{\'e}~F Bissyand{\'e}}, \bibinfo{person}{Dongsun
  Kim}, \bibinfo{person}{Kui Liu}, \bibinfo{person}{Jacques Klein},
  \bibinfo{person}{Martin Monperrus}, {and} \bibinfo{person}{Yves~Le Traon}.}
  \bibinfo{year}{2019}\natexlab{}.
\newblock \showarticletitle{D\&C: A Divide-and-Conquer Approach to IR-based Bug
  Localization}.
\newblock \bibinfo{journal}{\emph{arXiv preprint arXiv:1902.02703}}
  (\bibinfo{year}{2019}).
\newblock


\bibitem[\protect\citeauthoryear{Lam, Nguyen, Nguyen, and Nguyen}{Lam
  et~al\mbox{.}}{2017}]%
        {lam2017bug}
\bibfield{author}{\bibinfo{person}{An~Ngoc Lam}, \bibinfo{person}{Anh~Tuan
  Nguyen}, \bibinfo{person}{Hoan~Anh Nguyen}, {and} \bibinfo{person}{Tien~N
  Nguyen}.} \bibinfo{year}{2017}\natexlab{}.
\newblock \showarticletitle{Bug localization with combination of deep learning
  and information retrieval}. In \bibinfo{booktitle}{\emph{2017 IEEE/ACM 25th
  International Conference on Program Comprehension (ICPC)}}. IEEE,
  \bibinfo{pages}{218--229}.
\newblock


\bibitem[\protect\citeauthoryear{Le, Thung, and Lo}{Le et~al\mbox{.}}{2017}]%
        {le2017will}
\bibfield{author}{\bibinfo{person}{Tien-Duy~B Le}, \bibinfo{person}{Ferdian
  Thung}, {and} \bibinfo{person}{David Lo}.} \bibinfo{year}{2017}\natexlab{}.
\newblock \showarticletitle{Will this localization tool be effective for this
  bug? Mitigating the impact of unreliability of information retrieval based
  bug localization tools}.
\newblock \bibinfo{journal}{\emph{Empirical Software Engineering}}
  \bibinfo{volume}{22}, \bibinfo{number}{4} (\bibinfo{year}{2017}),
  \bibinfo{pages}{2237--2279}.
\newblock


\bibitem[\protect\citeauthoryear{Lee, Kim, Bissyand{\'e}, Jung, and
  Le~Traon}{Lee et~al\mbox{.}}{2018}]%
        {lee2018bench4bl}
\bibfield{author}{\bibinfo{person}{Jaekwon Lee}, \bibinfo{person}{Dongsun Kim},
  \bibinfo{person}{Tegawend{\'e}~F Bissyand{\'e}}, \bibinfo{person}{Woosung
  Jung}, {and} \bibinfo{person}{Yves Le~Traon}.}
  \bibinfo{year}{2018}\natexlab{}.
\newblock \showarticletitle{Bench4bl: reproducibility study on the performance
  of ir-based bug localization}. In \bibinfo{booktitle}{\emph{Proceedings of
  the 27th ACM SIGSOFT International Symposium on Software Testing and
  Analysis}}. \bibinfo{pages}{61--72}.
\newblock


\bibitem[\protect\citeauthoryear{Liang, Sun, Wang, and Yang}{Liang
  et~al\mbox{.}}{2019}]%
        {liang2019deep}
\bibfield{author}{\bibinfo{person}{Hongliang Liang}, \bibinfo{person}{Lu Sun},
  \bibinfo{person}{Meilin Wang}, {and} \bibinfo{person}{Yuxing Yang}.}
  \bibinfo{year}{2019}\natexlab{}.
\newblock \showarticletitle{Deep Learning With Customized Abstract Syntax Tree
  for Bug Localization}.
\newblock \bibinfo{journal}{\emph{IEEE Access}}  \bibinfo{volume}{7}
  (\bibinfo{year}{2019}), \bibinfo{pages}{116309--116320}.
\newblock


\bibitem[\protect\citeauthoryear{Lukins, Kraft, and Etzkorn}{Lukins
  et~al\mbox{.}}{2010}]%
        {lukins2010bug}
\bibfield{author}{\bibinfo{person}{Stacy~K Lukins}, \bibinfo{person}{Nicholas~A
  Kraft}, {and} \bibinfo{person}{Letha~H Etzkorn}.}
  \bibinfo{year}{2010}\natexlab{}.
\newblock \showarticletitle{Bug localization using latent dirichlet
  allocation}.
\newblock \bibinfo{journal}{\emph{Information and Software Technology}}
  \bibinfo{volume}{52}, \bibinfo{number}{9} (\bibinfo{year}{2010}),
  \bibinfo{pages}{972--990}.
\newblock


\bibitem[\protect\citeauthoryear{Markovtsev and Long}{Markovtsev and
  Long}{2018}]%
        {markovtsev2018public}
\bibfield{author}{\bibinfo{person}{Vadim Markovtsev} {and}
  \bibinfo{person}{Waren Long}.} \bibinfo{year}{2018}\natexlab{}.
\newblock \showarticletitle{Public Git archive: A big code dataset for all}. In
  \bibinfo{booktitle}{\emph{Proceedings of the 15th International Conference on
  Mining Software Repositories}}. \bibinfo{pages}{34--37}.
\newblock


\bibitem[\protect\citeauthoryear{Pietri, Spinellis, and Zacchiroli}{Pietri
  et~al\mbox{.}}{2019}]%
        {pietri2019software}
\bibfield{author}{\bibinfo{person}{Antoine Pietri}, \bibinfo{person}{Diomidis
  Spinellis}, {and} \bibinfo{person}{Stefano Zacchiroli}.}
  \bibinfo{year}{2019}\natexlab{}.
\newblock \showarticletitle{The software heritage graph dataset: public
  software development under one roof}. In \bibinfo{booktitle}{\emph{2019
  IEEE/ACM 16th International Conference on Mining Software Repositories
  (MSR)}}. IEEE, \bibinfo{pages}{138--142}.
\newblock


\bibitem[\protect\citeauthoryear{Polisetty, Miranskyy, and
  Ba{\c{s}}ar}{Polisetty et~al\mbox{.}}{2019}]%
        {polisetty2019usefulness}
\bibfield{author}{\bibinfo{person}{Sravya Polisetty}, \bibinfo{person}{Andriy
  Miranskyy}, {and} \bibinfo{person}{Ay{\c{s}}e Ba{\c{s}}ar}.}
  \bibinfo{year}{2019}\natexlab{}.
\newblock \showarticletitle{On Usefulness of the Deep-Learning-Based Bug
  Localization Models to Practitioners}. In
  \bibinfo{booktitle}{\emph{Proceedings of the Fifteenth International
  Conference on Predictive Models and Data Analytics in Software Engineering}}.
  \bibinfo{pages}{16--25}.
\newblock


\bibitem[\protect\citeauthoryear{Rath and M{\"a}der}{Rath and
  M{\"a}der}{2019}]%
        {rath2019structured}
\bibfield{author}{\bibinfo{person}{Michael Rath} {and} \bibinfo{person}{Patrick
  M{\"a}der}.} \bibinfo{year}{2019}\natexlab{}.
\newblock \showarticletitle{Structured information in bug report
  descriptions—influence on IR-based bug localization and developers}.
\newblock \bibinfo{journal}{\emph{Software Quality Journal}}
  \bibinfo{volume}{27}, \bibinfo{number}{3} (\bibinfo{year}{2019}),
  \bibinfo{pages}{1315--1337}.
\newblock


\bibitem[\protect\citeauthoryear{Rath, Rempel, and M{\"a}der}{Rath
  et~al\mbox{.}}{2017}]%
        {rath2017ilmseven}
\bibfield{author}{\bibinfo{person}{Michael Rath}, \bibinfo{person}{Patrick
  Rempel}, {and} \bibinfo{person}{Patrick M{\"a}der}.}
  \bibinfo{year}{2017}\natexlab{}.
\newblock \showarticletitle{The IlmSeven Dataset}. In
  \bibinfo{booktitle}{\emph{2017 IEEE 25th International Requirements
  Engineering Conference (RE)}}. IEEE, \bibinfo{pages}{516--519}.
\newblock


\bibitem[\protect\citeauthoryear{Saha, Lease, Khurshid, and Perry}{Saha
  et~al\mbox{.}}{2013}]%
        {saha2013improving}
\bibfield{author}{\bibinfo{person}{Ripon~K Saha}, \bibinfo{person}{Matthew
  Lease}, \bibinfo{person}{Sarfraz Khurshid}, {and} \bibinfo{person}{Dewayne~E
  Perry}.} \bibinfo{year}{2013}\natexlab{}.
\newblock \showarticletitle{Improving bug localization using structured
  information retrieval}. In \bibinfo{booktitle}{\emph{2013 28th IEEE/ACM
  International Conference on Automated Software Engineering (ASE)}}. IEEE,
  \bibinfo{pages}{345--355}.
\newblock


\bibitem[\protect\citeauthoryear{Wang and Lo}{Wang and Lo}{2014}]%
        {wang2014version}
\bibfield{author}{\bibinfo{person}{Shaowei Wang} {and} \bibinfo{person}{David
  Lo}.} \bibinfo{year}{2014}\natexlab{}.
\newblock \showarticletitle{Version history, similar report, and structure:
  Putting them together for improved bug localization}. In
  \bibinfo{booktitle}{\emph{Proceedings of the 22nd International Conference on
  Program Comprehension}}. \bibinfo{pages}{53--63}.
\newblock


\bibitem[\protect\citeauthoryear{Wong, Xiong, Zhang, Hao, Zhang, and Mei}{Wong
  et~al\mbox{.}}{2014}]%
        {wong2014boosting}
\bibfield{author}{\bibinfo{person}{Chu-Pan Wong}, \bibinfo{person}{Yingfei
  Xiong}, \bibinfo{person}{Hongyu Zhang}, \bibinfo{person}{Dan Hao},
  \bibinfo{person}{Lu Zhang}, {and} \bibinfo{person}{Hong Mei}.}
  \bibinfo{year}{2014}\natexlab{}.
\newblock \showarticletitle{Boosting bug-report-oriented fault localization
  with segmentation and stack-trace analysis}. In
  \bibinfo{booktitle}{\emph{2014 IEEE International Conference on Software
  Maintenance and Evolution}}. IEEE, \bibinfo{pages}{181--190}.
\newblock


\bibitem[\protect\citeauthoryear{Wong, Gao, Li, Abreu, and Wotawa}{Wong
  et~al\mbox{.}}{2016}]%
        {wong2016survey}
\bibfield{author}{\bibinfo{person}{W~Eric Wong}, \bibinfo{person}{Ruizhi Gao},
  \bibinfo{person}{Yihao Li}, \bibinfo{person}{Rui Abreu}, {and}
  \bibinfo{person}{Franz Wotawa}.} \bibinfo{year}{2016}\natexlab{}.
\newblock \showarticletitle{A survey on software fault localization}.
\newblock \bibinfo{journal}{\emph{IEEE Transactions on Software Engineering}}
  \bibinfo{volume}{42}, \bibinfo{number}{8} (\bibinfo{year}{2016}),
  \bibinfo{pages}{707--740}.
\newblock


\bibitem[\protect\citeauthoryear{Xiao, Keung, Mi, and Bennin}{Xiao
  et~al\mbox{.}}{2017}]%
        {xiao2017improving}
\bibfield{author}{\bibinfo{person}{Yan Xiao}, \bibinfo{person}{Jacky Keung},
  \bibinfo{person}{Qing Mi}, {and} \bibinfo{person}{Kwabena~E Bennin}.}
  \bibinfo{year}{2017}\natexlab{}.
\newblock \showarticletitle{Improving bug localization with an enhanced
  convolutional neural network}. In \bibinfo{booktitle}{\emph{2017 24th
  Asia-Pacific Software Engineering Conference (APSEC)}}. IEEE,
  \bibinfo{pages}{338--347}.
\newblock


\bibitem[\protect\citeauthoryear{Youm, Ahn, Kim, and Lee}{Youm
  et~al\mbox{.}}{2015}]%
        {youm2015bug}
\bibfield{author}{\bibinfo{person}{Klaus~Changsun Youm}, \bibinfo{person}{June
  Ahn}, \bibinfo{person}{Jeongho Kim}, {and} \bibinfo{person}{Eunseok Lee}.}
  \bibinfo{year}{2015}\natexlab{}.
\newblock \showarticletitle{Bug localization based on code change histories and
  bug reports}. In \bibinfo{booktitle}{\emph{2015 Asia-Pacific Software
  Engineering Conference (APSEC)}}. IEEE, \bibinfo{pages}{190--197}.
\newblock


\bibitem[\protect\citeauthoryear{Yu, Li, Yin, Wang, and Wang}{Yu
  et~al\mbox{.}}{2018}]%
        {yu2018dataset}
\bibfield{author}{\bibinfo{person}{Yue Yu}, \bibinfo{person}{Zhixing Li},
  \bibinfo{person}{Gang Yin}, \bibinfo{person}{Tao Wang}, {and}
  \bibinfo{person}{Huaimin Wang}.} \bibinfo{year}{2018}\natexlab{}.
\newblock \showarticletitle{A dataset of duplicate pull-requests in github}. In
  \bibinfo{booktitle}{\emph{Proceedings of the 15th International Conference on
  Mining Software Repositories}}. \bibinfo{pages}{22--25}.
\newblock


\bibitem[\protect\citeauthoryear{Zhou, Zhang, and Lo}{Zhou
  et~al\mbox{.}}{2012}]%
        {zhou2012should}
\bibfield{author}{\bibinfo{person}{Jian Zhou}, \bibinfo{person}{Hongyu Zhang},
  {and} \bibinfo{person}{David Lo}.} \bibinfo{year}{2012}\natexlab{}.
\newblock \showarticletitle{Where should the bugs be fixed? more accurate
  information retrieval-based bug localization based on bug reports}. In
  \bibinfo{booktitle}{\emph{2012 34th International Conference on Software
  Engineering (ICSE)}}. IEEE, \bibinfo{pages}{14--24}.
\newblock


\end{thebibliography}


\end{document}